\documentclass[twocolumn,secnumarabic,amssymb, nobibnotes, aps, prd]{revtex4-1}
\pdfoutput=1 
\usepackage[T1]{fontenc} 
\usepackage[utf8]{inputenc}
\usepackage{amsfonts}    
\usepackage{graphics}      
\usepackage{epsfig}     
\usepackage[english]{babel} 
\usepackage{caption} 
\usepackage{subcaption} 
\usepackage{amsmath}
\usepackage{bm} 
\usepackage{multirow} 
\usepackage{array,booktabs} 
\usepackage{calc} 
\usepackage[table]{xcolor} 
\usepackage{color}   
\usepackage{makecell} 
 
\definecolor{aqua}{rgb}{0.0, 1.0, 1.0}
\definecolor{babyblue}{rgb}{0.54, 0.81, 0.94}
\definecolor{beaublue}{rgb}{0.74, 0.83, 0.9}
\definecolor{blizzardblue}{rgb}{0.93, 0.93, 0.93} 
\definecolor{cyan}{rgb}{0.0, 1.0, 1.0}
\newcolumntype{?}{!{\vrule width 0.8pt}} 

\def\tanb{\tan\beta}  
\def\hpm{H^\pm} 
\def\mhpm{m_{\hpm}}
\def\mh{m_h}
\def\mH{m_H}
\def\ma{m_A} 
\def\tb{\tan\beta}

\def\bma{\beta-\alpha}  

\newcommand{\RN}[1]{ 
  \textup{\uppercase\expandafter{\romannumeral#1}}%
\renewcommand\thesubfigure{(\alph{subfigure})} 
\captionsetup[sub]{labelformat=simple} 
}
  
\begin{document}   

\title{\boldmath Search for neutral Higgs bosons decaying to $b\bar{b}$ in the flipped 2HDM\\ at future $e^-e^+$ linear colliders} 
\author{Majid Hashemi } 
\email{majid.hashemi@cern.ch}  
\author{Gholamhossein Haghighat}
\email{hosseinhaqiqat$@$gmail.com}  
\affiliation{Physics Department, College of Sciences, Shiraz University, \\ Shiraz, 71946-84795, Iran}

\begin{abstract}
In this study, assuming the Type-Y (flipped) 2HDM at the SM-like scenario as the theoretical framework, observability of the additional heavy neutral Higgs bosons $H$ and $A$ is investigated through the signal process chain $e^- e^+ \rightarrow AH \rightarrow b\bar{b}b\bar{b}$ at a linear collider operating at the center-of-mass energy of 1.5 TeV. The assumed signal process is highly motivated by the enhancements in the $A/H\rightarrow b\bar{b}$ decays at relatively high $\tb$ values. Such enhancements result in the dominance of the mentioned decay modes even for Higgs masses above the threshold of the on-shell top quark pair production. Taking advantage of such a unique feature, several benchmark scenarios are studied. Simulating the detector response based on the SiD detector at the ILC, simulated events are analyzed to reconstruct the $H$ and $A$ Higgs bosons. The top quark pair production and $Z/\gamma$ production are the main SM background processes and are well under control. Results indicate that, the $H$ and $A$ Higgs bosons are observable with signals exceeding $5\sigma$ with possibility of mass measurement in all the tested scenarios. Specifically, the parameter space region enclosed with the mass ranges $m_H$=150-500 GeV and $m_A$=230-580 GeV with the $A/H$ mass splitting of 80 GeV is observable at the integrated luminosity of 500 $fb^{-1}$.  
\end{abstract}   
 
\maketitle 
\flushbottom  
 
\section*{Introduction}    
The Standard Model (SM) of elementary particles provided significant predictions which have been successfully verified by plenty of experimental observations. Since the existence of the Higgs boson, as one of the most considerable predictions of the SM, was experimentally confirmed \cite{HiggsObservationCMS,HiggsObservationATLAS}, much effort has been devoted to developing the extended versions of the SM. Such extensions are mainly motivated by the SM inability to explain the universe baryon asymmetry \cite{Trodden}, supersymmetry \cite{MSSM1}, axion models \cite{KIM1}, etc.
Extending various aspects of the SM, different kinds of extensions with different characteristics can be obtained. The simplest scalar structure, a single scalar doublet, was assumed in the SM leading to the prediction of a single Higgs boson \cite{Englert1,Higgs1,Higgs2,Kibble1,Higgs3,Kibble2}. Extending the scalar structure by adding another scalar doublet, the two-Higgs-doublet model (2HDM) \cite{2hdm_TheoryPheno,2hdm1,2hdm2,2hdm3,2hdm4_CompositeHiggs,2hdm_HiggsSector1,2hdm_HiggsSector2,Campos:2017dgc} is obtained. As one of the important consequences of using two scalar doublets, the 2HDM predicts the existence of four additional Higgs bosons. To be specific, five Higgs bosons including a light scalar $h$, a heavy scalar $H$, a pseudoscalar $A$ and two charged $H^{\pm}$ Higgs bosons are offered by the 2HDM. To respect experimental observations, one may assume that the light scalar Higgs boson $h$ predicted in the 2HDM is the SM-like Higgs boson. Therefore, the 2HDM features four yet undiscovered Higgs bosons discovery of which may help confirm the 2HDM. This study is aimed to investigate observability of the two additional neutral Higgs bosons $A$ and $H$ in the 2HDM at a linear collider.

A general 2HDM predicts tree level flavor-changing neutral currents (FCNC) which are suppressed in the SM and are strongly constrained by experiments. Ensuring natural flavor conservation in the 2HDM is however possible with the help of special scenarios of Higgs-fermion couplings. Such selective couplings can be derived from imposing the discrete $Z_2$ symmetry. It has been shown that there are four coupling scenarios permitted by the $Z_2$ symmetry which avoid tree level FCNCs \cite{2hdm_HiggsSector2}. Consequently, there are four types of the 2HDM with different phenomenologies which naturally conserve flavor. Observability of the two additional neutral Higgs bosons within the Type-\RN{1} and Type-X 2HDMs has been studied with promising results \cite{MHashemiMMahdavi-1-4b,H-2HDMX}.

This study considers the Type-Y (flipped) 2HDM and investigates the observability of the $H$ and $A$ Higgs bosons through the signal process chain $e^- e^+ \rightarrow AH \rightarrow b\bar{b}b\bar{b}$ where $b$ is the bottom quark. The assumed signal process is mainly motivated by the enhancements in the $A/H\rightarrow b\bar{b}$ decays at high $\tb$ values. Such enhancements as well as the suppression of the $A/H$ decay into a pair of up-type quarks at high values of $\tb$ result in the dominance of the $A/H\rightarrow b\bar{b}$ decays even for Higgs masses above the threshold of the on-shell top quark pair production. Consequently, a significantly large portion of the parameter space can be probed with the help of the considered signal process. This is a unique feature of the assumed signal process in the flipped 2HDM.

Assuming several benchmark points with different mass hypotheses, observability of the Higgs bosons is assessed by analyzing simulated events for each scenario independently. Because of the chosen Higgs mass ranges and the assumed signal process, the present analysis is most suitable for a collider experiment performed at the center-of-mass energy of 1.5 TeV. Although such an experiment can be easily performed by the LHC, a linear collider is assumed in this study since $e^-e^+$ linear colliders suffer less from background processes, underlying events, etc. Assuming both beams to be unpolarised, signal and background events are generated at the integrated luminosity of 500 $fb^{-1}$ and the detector response is simulated based on the SiD detector at the International Linear Collider (ILC) \cite{SiDatILC}. Reconstructing and identifying $b$-jets with the use of proper jet clustering and $b$-tagging algorithms, simulated events are analyzed to reconstruct the Higgs bosons. Computing invariant masses of the $b$ quark pairs coming from the Higgs bosons, we try to obtain a Higgs candidate mass distribution for each scenario. It will be shown that both of the $H$ and $A$ Higgs bosons are observable with signals exceeding $5\sigma$ with possibility of mass measurement in all the considered scenarios. To be specific, the region of parameter space enclosed with the mass ranges $150\leq m_H \leq 500$ GeV and $230\leq m_A \leq 580$ GeV with the $A/H$ mass splitting of 80 GeV is observable at the integrated luminosity of 500 $fb^{-1}$. In what follows, we present a brief introduction to the 2HDM and then different aspects of the analysis will be discussed.

\section{Theoretical framework} 
Extending the Standard Model by adding another $SU(2)$ Higgs doublet and postulating the general Higgs potential
\begin{equation}
  \begin{aligned}
     \mathcal{V} & =\, m_{11}^2\Phi_1^\dagger\Phi_1+m_{22}^2\Phi_2^\dagger\Phi_2
    -\Big[m_{12}^2\Phi_1^\dagger\Phi_2+\mathrm{h.c.}\Big]
    \\
    &+\frac{1}{2}\lambda_1\Big(\Phi_1^\dagger\Phi_1\Big)^2
    +\frac{1}{2}\lambda_2\Big(\Phi_2^\dagger\Phi_2\Big)^2
    +\lambda_3\Big(\Phi_1^\dagger\Phi_1\Big)\Big(\Phi_2^\dagger\Phi_2\Big)
    \\& +\lambda_4\Big(\Phi_1^\dagger\Phi_2\Big)\Big(\Phi_2^\dagger\Phi_1\Big)
    +\Big\{\frac{1}{2}\lambda_5\Big(\Phi_1^\dagger\Phi_2\Big)^2
    +\Big[\lambda_6\Big(\Phi_1^\dagger\Phi_1\Big)
     \\& +\lambda_7\Big(\Phi_2^\dagger\Phi_2\Big)
      \Big]\Big(\Phi_1^\dagger\Phi_2\Big)
    +\mathrm{h.c.}\Big\},
  \end{aligned}
  \label{lag}
\end{equation}
where $\Phi_1$ and $\Phi_2$ are $SU(2)$ Higgs doublets, one of the simplest extensions of the SM, the 2HDM \cite{2hdm_TheoryPheno,2hdm1,2hdm2,2hdm3,2hdm4_CompositeHiggs,2hdm_HiggsSector1,2hdm_HiggsSector2,Campos:2017dgc}, is obtained. The two assumed Higgs doublets have eight degrees of freedom, three of which are ``eaten'' by three of the electroweak gauge bosons $W^\pm, Z$ and the remaining five degrees of freedom lead to the prediction of five Higgs bosons, namely the neutral light $h$ and heavy $H$ scalar, the neutral pseudoscalar $A$ and the charged $H^\pm$ Higgs bosons. To completely specify the model, the parameters $\tanb$, $m_{12}^2$, $\lambda_6$, $\lambda_7$, mixing angle $\alpha$ and physical Higgs masses $m_h, m_H, m_A, m_{H^\pm}$ must be determined in the ``physical basis'' \cite{2hdm_TheoryPheno}.

A general 2HDM gives rise to flavor-changing neutral currents (FCNC) at tree level which are absent in the SM and are strongly constrained by experimental observations. Introducing the $Z_2$ symmetry, such currents are well avoided in the scalar sector and models with natural flavor conservation are obtained \cite{2hdm2,2hdm3,2hdm4_CompositeHiggs}. The imposed $Z_2$ symmetry implies that the Higgs coupling to fermions must follow the scenarios shown in Tab. \ref{coupling}. 
\begin{table}[h]
\normalsize
\fontsize{11}{7.2} 
    \begin{center}
         \begin{tabular}{ >{\centering\arraybackslash}m{.8in} >{\centering\arraybackslash}m{.4in} >{\centering\arraybackslash}m{.4in} >{\centering\arraybackslash}m{.4in}   }
& {$u_R^i$} & {$d_R^i$} & {$\ell_R^i$} \parbox{0pt}{\rule{0pt}{1ex+\baselineskip}}\\ \Xhline{3\arrayrulewidth}
 {Type \RN{1}} &$\Phi_2$ &$\Phi_2$ &$\Phi_2$ \parbox{0pt}{\rule{0pt}{1ex+\baselineskip}}\\ 
 {Type $\RN{2}$} &$\Phi_2$ &$\Phi_1$ &$\Phi_1$ \parbox{0pt}{\rule{0pt}{1ex+\baselineskip}}\\ 
{Type X} &$\Phi_2$ &$\Phi_2$ &$\Phi_1$  \parbox{0pt}{\rule{0pt}{1ex+\baselineskip}}\\ 
{Type Y} &$\Phi_2$ &$\Phi_1$ &$\Phi_2$  \parbox{0pt}{\rule{0pt}{1ex+\baselineskip}}\\ \Xhline{3\arrayrulewidth}
  \end{tabular}
\caption{Higgs coupling to up-type quarks, down-type quarks and leptons in different types of 2HDM. The superscript $i$ is a generation index.}
\label{coupling}
  \end{center}
\end{table}
As seen, there are four types of 2HDM which naturally conserve flavor. The types ``X'' and ``Y'' are also called ``lepton-specific'' and ``flipped'' respectively. As a consequence of the imposed $Z_2$ symmetry, the parameters $m_{12}^2$, $\lambda_6$ and $\lambda_7$ must be zero. However, allowing a non-zero value for $m_{12}^2$, $Z_2$ symmetry is softly broken. The parameters $m_{11}^2$ and $m_{22}^2$ in the Higgs potential relate to $\tb$ through minimization conditions for a minimum of the vacuum and can be obtained once $\tb$ is determined. 

To respect experimental observations, one can assume that the lightest scalar Higgs boson $h$ predicted in the 2HDM is the same as the observed SM Higgs boson. To do so, $h$ couplings to fermions in the Yukawa Lagrangian of the 2HDM must reduce to those of the SM. These selective couplings are easily implemented in a natural way through the SM-like assumption $\sin(\bma)=1$ \cite{2hdm_TheoryPheno}. Following the coupling scenarios provided in Tab. \ref{coupling} and applying the SM-like assumption, the neutral Higgs part of the Yukawa Lagrangian takes the form \cite{Barger_2hdmTypes,2hdm_TheoryPheno}
\begin{equation}
\begin{aligned}
& \mathcal{L}_{ Yukawa}\, =\,  -v^{-1}  \Big(\, m_d\, \bar{d}d\, +\, m_u\, \bar{u}u\, +\, m_\ell\, \bar{\ell}\ell\, \Big)\ h \\
      & +v^{-1} \Big(\, \rho^dm_d\, \bar{d}d\, +\, \rho^um_u\, \bar{u}u\, +\, \rho^\ell m_\ell\, \bar{\ell}\ell\, \Big)\, H \\
& +iv^{-1}\Big(-\rho^dm_d\, \bar{d}\gamma_5d\, +\, \rho^um_u\, \bar{u}\gamma_5u\, -\, \rho^\ell m_\ell\, \bar{\ell}\gamma_5\ell\, \Big) A,
\label{yukawa2}
\end{aligned}
\end{equation}
 where $\rho^X$ factors corresponding to different types are provided in Tab. \ref{rho}.
\begin{table}[h]
\normalsize
\fontsize{11}{7.2} 
    \begin{center}
         \begin{tabular}{ >{\centering\arraybackslash}m{.5in}  >{\centering\arraybackslash}m{.55in}  >{\centering\arraybackslash}m{.55in} >{\centering\arraybackslash}m{.55in}  >{\centering\arraybackslash}m{.55in}}
& {\RN{1}} & {$\RN{2}$} & {X} & {Y} \parbox{0pt}{\rule{0pt}{1ex+\baselineskip}}\\ \Xhline{3\arrayrulewidth}
 {$\rho^d$} &$\cot{\beta}$ &$- \tan\beta$ &$\cot\beta$ &$-\tan\beta$ \parbox{0pt}{\rule{0pt}{1ex+\baselineskip}}\\ 
{$\rho^u$} &$\cot{\beta}$ &$\cot\beta$ &$\cot\beta$ &$\cot\beta$  \parbox{0pt}{\rule{0pt}{1ex+\baselineskip}}\\   
{$\rho^\ell$} &$\cot{\beta}$ &$- \tan\beta$ &$-\tan\beta$&$\cot\beta$  \parbox{0pt}{\rule{0pt}{1ex+\baselineskip}}\\ \Xhline{3\arrayrulewidth} 
 \end{tabular}
\caption{$\rho^X$ factors in the neutral Higgs sector of the Yukawa Lagrangian in different types of 2HDM. } 
\label{rho}
  \end{center}          
\end{table}    
As seen, couplings are different in different types leading to dramatically different environments and phenomenologies \cite{2hdm_HiggsSector2}. According to Tab. \ref{rho}, Higgs coupling to down-type quarks depends on $-\tb$ in Type-Y. Consequently, annihilation of the $H$ and $A$ Higgs bosons into a pair of down-type quarks receives significant enhancements at high $\tb$ values. The present study takes advantage of such a feature and investigates observability of the $H$ and $A$ Higgs bosons in the framework of the Type-Y 2HDM at SM-like scenario.
 
\section{Signal process}
Observability of the additional neutral Higgs bosons within the flipped 2HDM is investigated through the signal production process $e^- e^+ \rightarrow AH$ with subsequent decays of the Higgs bosons into $b\bar{b}$ pairs where $b$ is the $b$ quark. The initial collision is assumed to occur at a linear collider operating at the center-of-mass energy of 1.5 TeV and the integrated luminosity is assumed to be 500 $fb^{-1}$. The considered signal process benefits from enhancements in the decay modes $A\rightarrow b\bar{b}$ and $H\rightarrow b\bar{b}$ which are due to the dependence of the $A/H$ coupling to down-type quarks on the $-\tb$ factor according to the Yukawa Lagrangian of Eq. \ref{yukawa2} and factors of Tab. \ref{rho}. Such a coupling factor results in dramatic enhancements and thus dominance of the $A/H\rightarrow b\bar{b}$ decays at relatively large values of $\tb$. Surprisingly, the dominance of these decay modes continues even for Higgs masses $m_{A/H}$ above the threshold of the on-shell top quark pair production. Such a feature is caused by the dependence of the $A/H$-$u$-$\bar{u}$ vertex, where $u$ is an up-type quark, on the $\cot\beta$ factor as seen in Tab. \ref{rho}. As $\tb$ increases, decays into the $t\bar{t}$ pair are becoming more and more rare and the $b\bar{b}$ pair production remains dominant. This is a unique feature of the assumed signal process in the context of the flipped 2HDM and enables us to probe a significantly large portion of the parameter space since our search is not limited to scenarios with Higgs masses below the threshold of the on-shell top quark pair production. 

Observability of the additional Higgs bosons is studied in several benchmark points in the parameter space of the 2HDM independently. Tab. \ref{BPsSignal} provides the assumed points with corresponding cross sections and branching fractions of the $A/H\rightarrow b\bar{b}$ decays. Working in the ``physical basis'', the assumed points are specified by physical Higgs masses, $m_{12}^2$, $\tb$ and $\sin(\bma)$. 
\begin {table*}[!htbp]  
\begin{subtable}[b]{.59\textwidth}
\centering
\begin{tabular}{cccccc} 
\multicolumn{6}{ c }{$\sqrt s=1.5$ TeV } \\ \Xhline{3\arrayrulewidth}
& BP1 & BP2 & BP3 & BP4 & BP5\\ \Xhline{3\arrayrulewidth}
$m_{h}$ & \multicolumn{5}{ c }{125} \\ 
$m_{H}$ & 150 & 200 & 300 & 400 & 500 \\
$m_{A}$ & 230 & 280 & 380 & 480 & 580 \\ 
$m_{H^\pm}$ & 230 & 280 & 380 & 480 & 580 \\ 
$m_{12}^2$ & 1093-1124 & 1966-1996  & 4459-4490 & 7951-7981 & 12439-12470 \\  
$\tan\beta$ & \multicolumn{5}{ c }{20} \\ 
$\sin(\beta-\alpha)$ & \multicolumn{5}{ c }{1} \\ 
$\sigma$ $[fb]$ & 5.7 & 5.2 & 4.1 & 2.9 & 1.7 \\ 
$BR(A\rightarrow b\bar{b})$ & 0.997 & 0.998 & 0.982 & 0.976 & 0.974 \\
$BR(H\rightarrow b\bar{b})$ & 0.998 & 0.999 & 0.999 & 0.994 & 0.986 \\ 
\Xhline{3\arrayrulewidth}
\end{tabular}
\caption {} 
\label{BPsSignal} 
\end{subtable}   
\begin{subtable}[b]{.4\textwidth}
\begin{center}     
\begin{tabular}{ccccc} 
$\sqrt s=1.5$ TeV & $t\bar{t}$ & $W^+W^-$ & $ZZ$ & $Z/\gamma$ \\ \Xhline{3\arrayrulewidth}
$\sigma$ $[fb]$ & 103 & 1796 & 131 & 1960  \\ \Xhline{3\arrayrulewidth}
\end{tabular}
\caption {}
\label{bgXsec}
\end{center} 
\end{subtable}  
\caption {a) Assumed benchmark scenarios. $\mh,\mH, \ma,\mhpm$ are physical masses of the Higgs bosons and the provided $m^2_{12}$ range satisfies the theoretical constraints. Cross section of the signal production process and branching fractions of the $A/H\rightarrow b\bar{b}$ decays are also provided for each scenario. b) Relevant SM background processes with corresponding cross sections.}
\label{BPs}
\end {table*} 
As seen, the mass of the additional CP-even Higgs boson is assumed to range from 150 to 500 GeV and the mass of the CP-odd Higgs boson $A$ is assumed to vary in range 230-580 GeV with the $A/H$ mass splitting of 80 GeV in all the scenarios. $\tb$ is set to 20 for all the scenarios for the signal to take advantage of the possible enhancements in the $A/H\rightarrow b\bar{b}$ decays at large values of $\tb$. According to the given branching fractions which are computed by \texttt{2HDMC 1.7.0} \cite{2hdmc1,2hdmc2}, on average, we have BR$(A \rightarrow b\bar{b})\simeq 0.985$ and BR$(H \rightarrow b\bar{b})\simeq 0.995$. Obviously, the $A/H\rightarrow b\bar{b}$ decays are dominant in all the scenarios. As seen, $\sin(\bma)$ is assumed to be 1 because of the SM-like assumption. The $h$ Higgs boson is therefore considered as the SM-like Higgs boson in all the scenarios. 

The assumed scenarios are all checked using \texttt{2HDMC 1.7.0} for consistency with theoretical constraints, namely potential stability \cite{Deshpande}, perturbativity and unitarity \cite{Huffel,Maalampi,KANEMURA,GAKEROYD} and the $m_{12}^2$ range satisfying the required constraints is provided in Tab. \ref{BPsSignal} for each scenario. 

As seen in Tab. \ref{BPsSignal}, the charged Higgs mass $m_{H^\pm}$ is chosen to be equal to the $H$ mass. The reason for making such a choice is that according to \cite{drho,Gerard:2007kn}, the deviation of the $\rho=m_W^2(m_Z\cos\theta_W)^{-2}$ parameter value in the 2HDM from its Standard Model value is negligible if any of the conditions
\begin{equation} 
m_A=m_{H^\pm},\,\,\, m_H=m_{H^\pm},
\label{negligibledrho}  
\end{equation}
is met. Hence, the assumed scenarios satisfy the strong experimental constraint \cite{BERTOLINI,DENNER} on the $\rho$ deviation which is based on the measurement performed at LEP \cite{Yao}. 

The LHC experiments \cite{ATLAS-2HDM-2} constrain the $A$ mass by the upper limits $m_A\leq 250,\, 295,\, 400,\, 510,\, 640$ GeV for the $H$ masses $m_H=150,\, 200,\, 300,\, 400,\, 500$ GeV respectively at $\tb=20$ in the Type-Y 2HDM. Obviously, the chosen scenarios satisfy these constraints and therefore, are safe to use.

In the context of the Type-\RN{1}, the lower limit $m_A>350$ obtained by the LHC direct observations  \cite{CMS-2HDM-2,ATLAS-2HDM-1} constrains the CP-odd Higgs mass for $\tb<5$. Also, the $H$ mass range 170-360 GeV has been excluded for $\tb<1.5$ \cite{ATLAS-2HDM-3}. However, since the Higgs-fermion coupling scenarios of the Type-\RN{1} and Type-Y are dramatically different ,Higgs masses in this study are not required to satisfy these limits.

In the context of the MSSM, the LEP experiments \cite{lep1,lep2,lepexclusion2} put the lower limits $m_A\geq93.4$ GeV and $m_{H^\pm}\geq78.6$ GeV on the $A$ and charged Higgs masses and the mass range $m_{A/H}=200-400$ GeV is also excluded for $\tb\geq5$ by the LHC experiments \cite{CMSNeutralHiggs,ATLASNeutralHiggs}. The experimental constraints on the MSSM are not, however, required to be satisfied by the assumed scenarios in this study since the MSSM and Type-Y 2HDM completely differ in many aspects, namely imposed symmetries, Higgs couplings, free parameters, etc. Consequently, it can be concluded that the assumed benchmark scenarios are completely consistent with all the theoretical and current experimental constraints.

Signal and background events are generated according to the assumed scenarios and the simulated detector response is analyzed to reconstruct the $H$ and $A$ Higgs bosons by finding $b\bar{b}$ pairs coming from their decays. $W^\pm$ pair production, $Z/\gamma$ production, $Z$ pair production and top quark pair production are the relevant SM background processes which are taken into account in this analysis. Cross sections of the signal and background processes are obtained by \texttt{PYTHIA 8.2.15} \cite{pythia82} and are provided in Tab. \ref{BPs}.      
    
\section{Event generation, signal selection and analysis}   
Assuming both beams to be unpolarised, basic parameters of the Type-Y 2HDM are produced in SLHA (SUSY Les Houches Accord) format by \texttt{2HDMC 1.7.0} and the output file is passed to \texttt{PYTHIA 8.2.15} \cite{pythia82} to generate events. Events generated by \texttt{PYTHIA} are internally used by \texttt{DELPHES 3.4} \cite{DELPHES3.4} to simulate the detector response with the use of DSiD detector card which is based on the full simulation performance of the SiD detector at the ILC \cite{SiDatILC}. Jet reconstruction is performed by the anti-$k_t$ algorithm \cite{antikt} in FASTJET 3.1.0 \cite{fastjet1,fastjet2} with the cone size $\Delta R=\sqrt{(\Delta\eta)^2+(\Delta\phi)^2}=0.4$, where $\eta=-\textnormal{ln}\tan(\theta/2)$ and $\phi$ ($\theta$) is the azimuthal (polar) angle with respect to the beam axis. The \texttt{DELPHES} output data including reconstructed jets and associated $b$-tagging flags are stored as \texttt{ROOT} files \cite{root1} and are analyzed as follows.

\begin{figure}[!htbp]
  \centering
    \begin{subfigure}[b]{0.48\textwidth}
    \centering 
    \includegraphics[width=\textwidth]{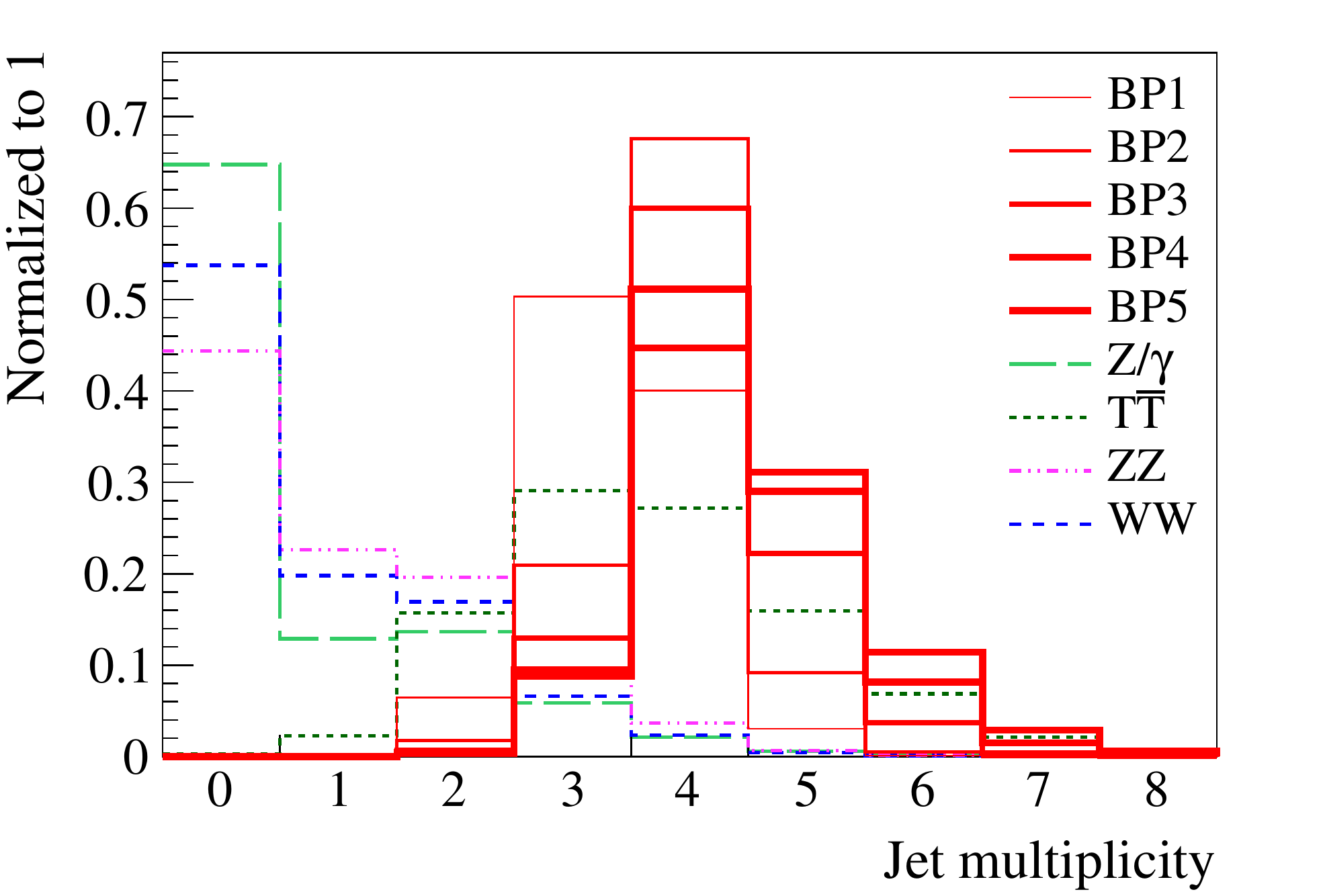}
    \caption{}
    \label{hnjets} 
    \end{subfigure}  
        \quad    
    \begin{subfigure}[b]{0.48\textwidth}
    \centering
    \includegraphics[width=\textwidth]{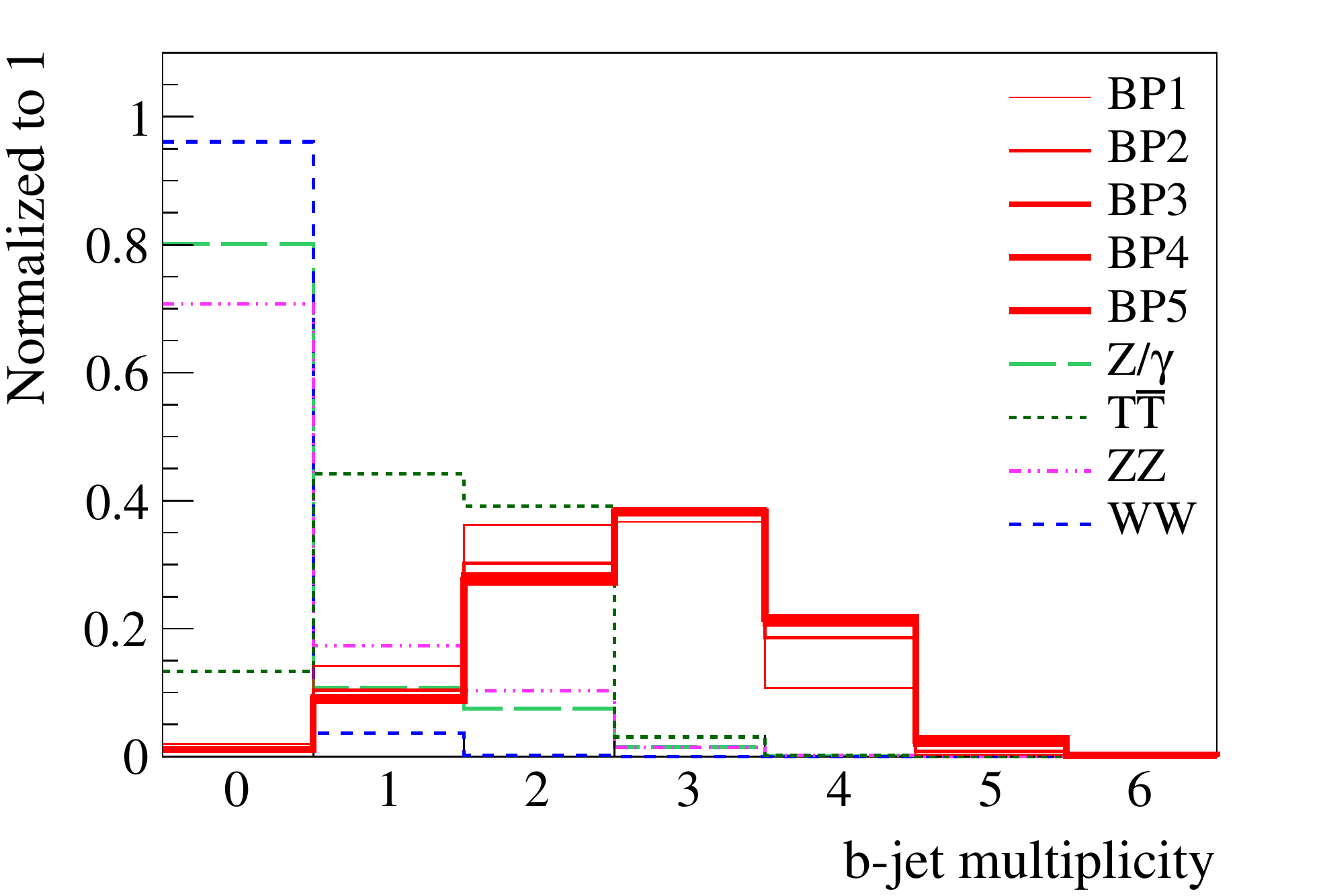}
    \caption{}
    \label{hnbjets}
    \end{subfigure}
  \caption{a) Jet and b) $b$-jet multiplicity distributions corresponding to different signal and background processes assuming different benchmark scenarios.}
  \label{hnjetshnbjetshndileptons} 
\end{figure}

Counting the reconstructed jets satisfying the kinematic conditions
\begin{equation}
\begin{aligned}
& \bm{{p_T}}_{\bm{jet}}\geq30\ GeV,\ \ \  \vert \bm{\eta}_{\bm{jet}} \vert \leq 2,  
\end{aligned}
\label{jetconditions} 
\end{equation}  
where $p_T$ is the transverse momentum, jet multiplicity distributions of Fig. \ref{hnjets} are obtained for different signal and background processes. 
Based on the signal and background distributions, the selection cut
\begin{equation}
\bm{N}_{\textbf{\emph{jet}}} \geq 3,
\label{jetnumbercondition} 
\end{equation}
where $N_{\textbf{\emph{jet}}}$ is the number of jets, is applied to events. Using the $b$-tagging flags, $b$-jets are identified and $b$-jet multiplicity distributions of Fig. \ref{hnbjets} are obtained. The condition 
\begin{equation}
\bm{N}_{\textbf{\emph{b-jet}}} \geq 3,
\label{bjetnumbercondition} 
\end{equation}
where $N_{\textbf{\emph{b-jet}}}$ is the number of $b$-jets, is then imposed and events surviving this condition are used to reconstruct the Higgs bosons. Applying the mentioned selection cuts, event selection efficiencies of Tab. \ref{eff} are obtained for different signal and background processes.
\begin{table}[!htbp] 
\begin{subtable}[b]{.48\textwidth}
\normalsize
\fontsize{11}{7.2} 
    \begin{center}
        \begin{tabular}{>{\centering\arraybackslash}m{.88in}  >{\centering\arraybackslash}m{.4in}  >{\centering\arraybackslash}m{.4in} >{\centering\arraybackslash}m{.4in} >{\centering\arraybackslash}m{.4in} >{\centering\arraybackslash}m{.44in}}
 \Xhline{3\arrayrulewidth}
  & {BP1} & {BP2} & {BP3} & {BP4} & {BP5} \parbox{0pt}{\rule{0pt}{1ex+\baselineskip}}\\ \Xhline{3\arrayrulewidth}
    {$N_{jet}\geq3$} & 0.935 & 0.982 & 0.993 & 0.996 & 0.996 \parbox{0pt}{\rule{0pt}{1ex+\baselineskip}}\\ 
   {$N_{\emph{b-jet}}\geq 3$} & 0.476 & 0.581 & 0.610 & 0.625 & 0.629 \parbox{0pt}{\rule{0pt}{1ex+\baselineskip}}\\ 
    {\textbf{Total eff.}} & \textbf{0.446} & \textbf{0.570} & \textbf{0.606} & \textbf{0.622} & \textbf{0.626} \parbox{0pt}{\rule{0pt}{1ex+\baselineskip}}\\ \Xhline{3\arrayrulewidth}
        \end{tabular}
\caption{} 
\label{signaleff}
  \end{center}
\end{subtable}
\newline\newline
\begin{subtable}[b]{.48\textwidth}
\normalsize
\fontsize{11}{7.2} 
    \begin{center}
        \begin{tabular}{ >{\centering\arraybackslash}m{.88in}  >{\centering\arraybackslash}m{.51in}  >{\centering\arraybackslash}m{.51in} >{\centering\arraybackslash}m{.51in} >{\centering\arraybackslash}m{.55in} }
\Xhline{3\arrayrulewidth}
  & {$t\bar{t}$} & {$WW$} & {$ZZ$} & {$Z/\gamma$ } \parbox{0pt}{\rule{0pt}{1ex+\baselineskip}}\\ \Xhline{3\arrayrulewidth}
    {$N_{jet}\geq3$} & 0.818 & 0.095 & 0.134 & 0.087  \parbox{0pt}{\rule{0pt}{1ex+\baselineskip}}\\ 
   {$N_{\emph{b-jet}}\geq 3$} & 0.033 & 1e-4 & 0.017 & 0.017 \parbox{0pt}{\rule{0pt}{1ex+\baselineskip}}\\ 
    {\textbf{Total eff.}} & \textbf{0.027} & \textbf{1e-05} & \textbf{0.002} & \textbf{0.001} \parbox{0pt}{\rule{0pt}{1ex+\baselineskip}}\\ \Xhline{3\arrayrulewidth}
        \end{tabular}
\caption{}
\label{bkgeff}
  \end{center}
\end{subtable}
 \caption{Event selection efficiencies obtained for the a) signal and b) background processes assuming different benchmark scenarios.}
\label{eff}
\end{table}

Events surviving the selection cuts contain at least three $b$-jets which are used to obtain the candidate mass distribution of the Higgs bosons. In events with three $b$-jets, $\Delta R_{\, bb}$, where $\Delta R=\sqrt{(\Delta\eta)^2+(\Delta\phi)^2}$, is computed for the three possible $bb$ combinations and the combination with minimum $\Delta R_{\, bb}$ is identified as the correct $b$-jet pair which originates from the decay of the $H$ or $A$ Higgs boson. In events with at least four $b$-jets, two pairs of $b$-jets coming from the Higgs bosons must be identified. To do so, the $b$-jets are sorted in terms of their energies. Labeling the sorted $b$-jets as $b_1, b_2, b_3, b_4$, the pairs $b_1b_4$ and $b_2b_3$ are considered as correct pairs. The first pair consists of the $b$-jets with lowest and highest energies and the second pair consists of the $b$-jets with moderate energies. Each one of these pairs may come from the $H$ or $A$ Higgs boson. Therefore, the distribution of the invariant masses of the identified $b$-jet pairs is expected to show two distinguished peaks since the Higgs bosons $H$ and $A$ are assumed to have different masses.

The analysis can be improved by applying a correction to the four-momentums of the $b$-jets based on the energy-momentum conservation. Assuming that the $b$-jet's flight direction has correctly been measured and a common factor can be applied to all components of its four-momentum, the linear system
\begin{align}
\begin{split}
z_1~p_1^x + z_2~p_2^x + z_3~p_3^x + z_4~p_4^x &= 0, \\  
z_1~p_1^y + z_2~p_2^y + z_3~p_3^y + z_4~p_4^y &= 0, \\ 
z_1~p_1^z + z_2~p_2^z + z_3~p_3^z + z_4~p_4^z &= 0, \\
z_1~E_1 + z_2~E_2 + z_3~E_3 + z_4~E_4 &= \sqrt{s},\\
\end{split}
\label{zfactors}
\end{align}
where $p_i^X$ and $E_i$ are the $X$-direction component of the three-momentum and the energy of the $i$'th $b$-jet respectively, is simultaneously solved to find the unknown variables $z_1, z_2, z_3, z_4$ which are factors corresponding to the four $b$-jets. Requiring the obtained factors to be positive (which is the case for a majority of the signal events) and applying the factors to the $b$-jets' four-momenta, a significant improvement in the invariant mass distributions is achieved. 

\begin{figure*}[h]
  \centering  
    \begin{subfigure}[b]{0.49\textwidth} 
    \centering
    \includegraphics[width=\textwidth]{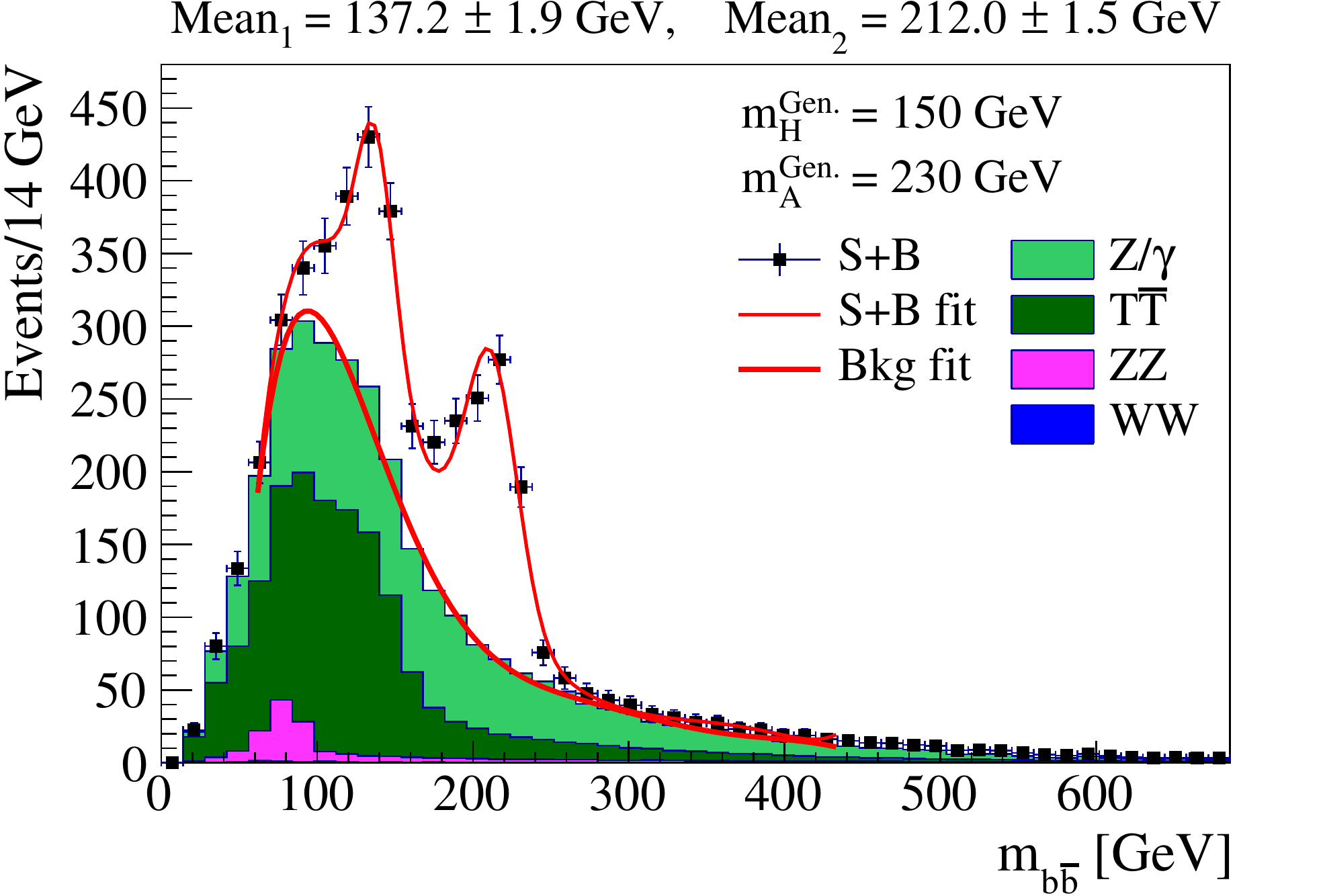}
    \caption{}
    \label{BP1}
    \end{subfigure} 
    \begin{subfigure}[b]{0.49\textwidth}
    \centering
    \includegraphics[width=\textwidth]{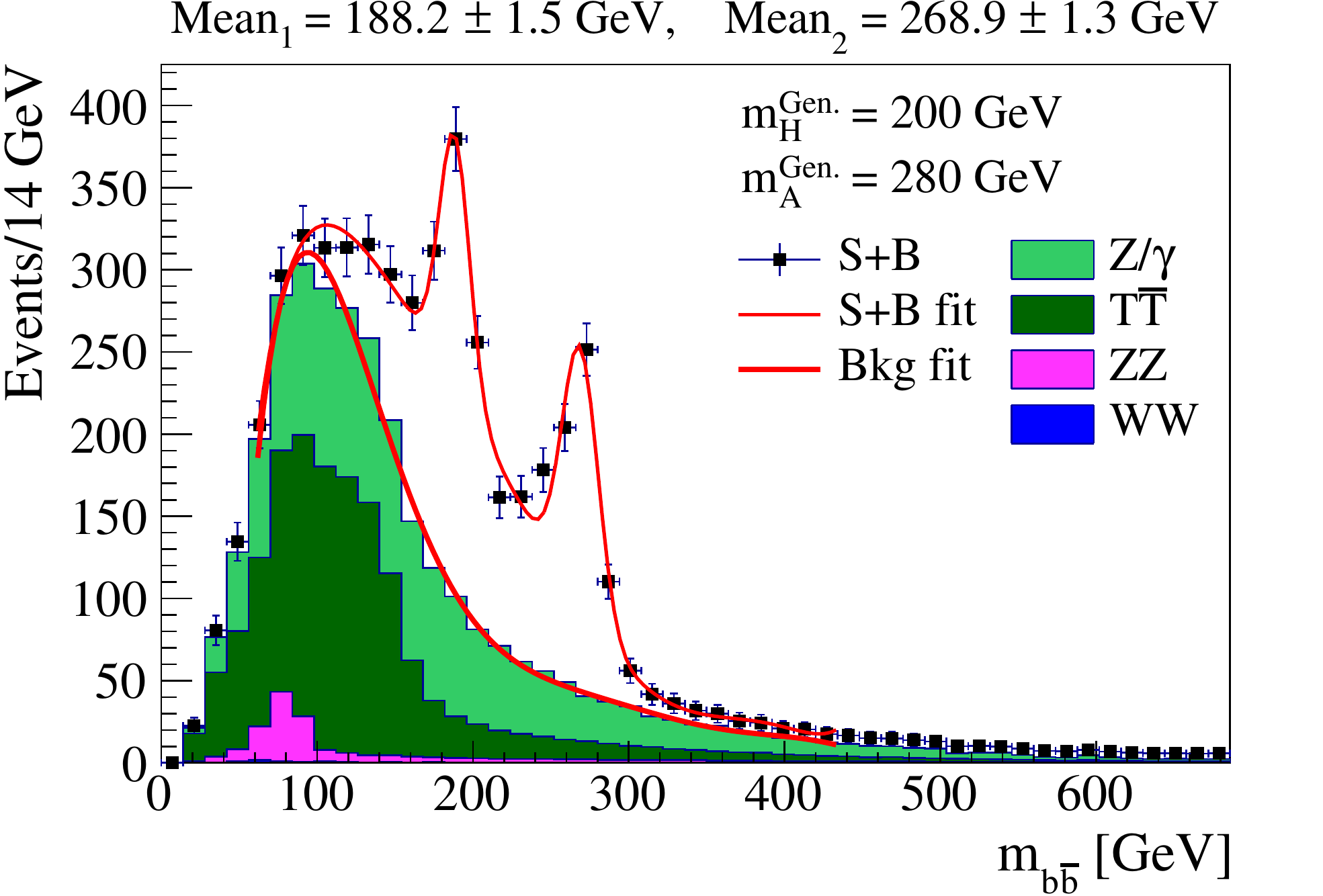}
    \caption{}
    \label{BP2} 
    \end{subfigure} 
\newline\newline
    \begin{subfigure}[b]{0.49\textwidth}
    \centering
    \includegraphics[width=\textwidth]{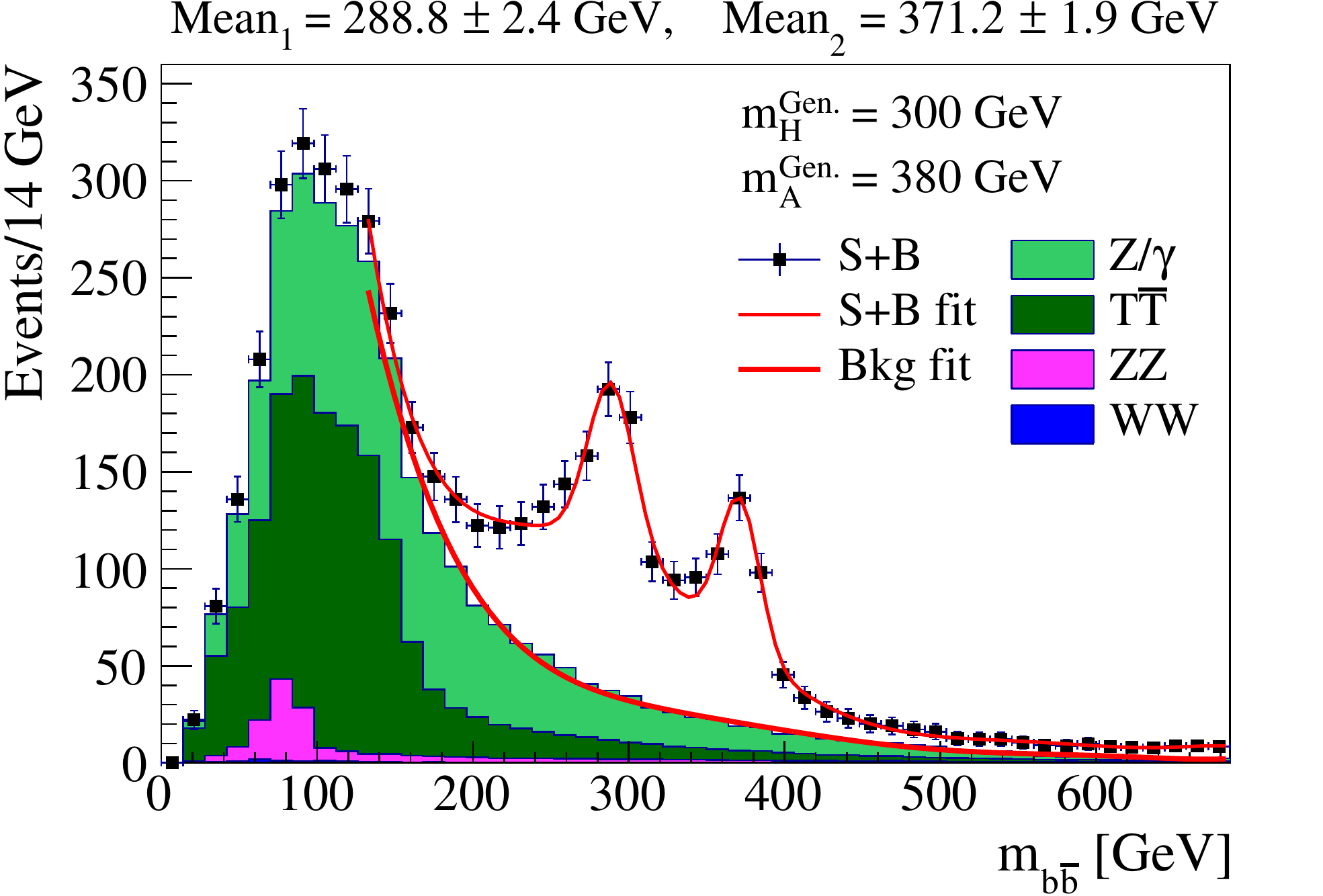}
    \caption{}
    \label{BP3}
    \end{subfigure}
    \begin{subfigure}[b]{0.49\textwidth}
    \centering
    \includegraphics[width=\textwidth]{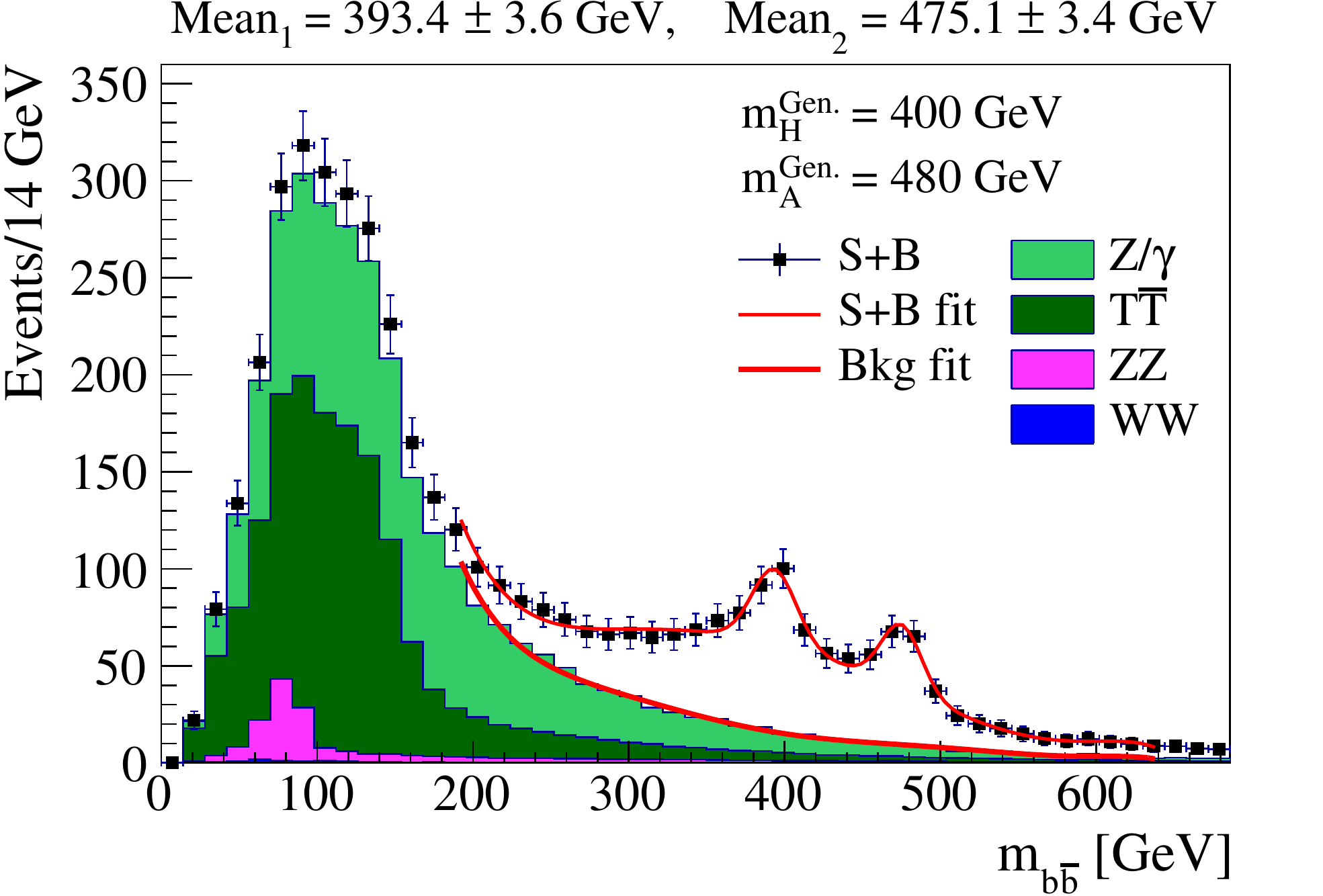}
    \caption{}
    \label{BP4} 
    \end{subfigure}
\newline\newline
    \begin{subfigure}[b]{0.49\textwidth} 
    \centering
    \includegraphics[width=\textwidth]{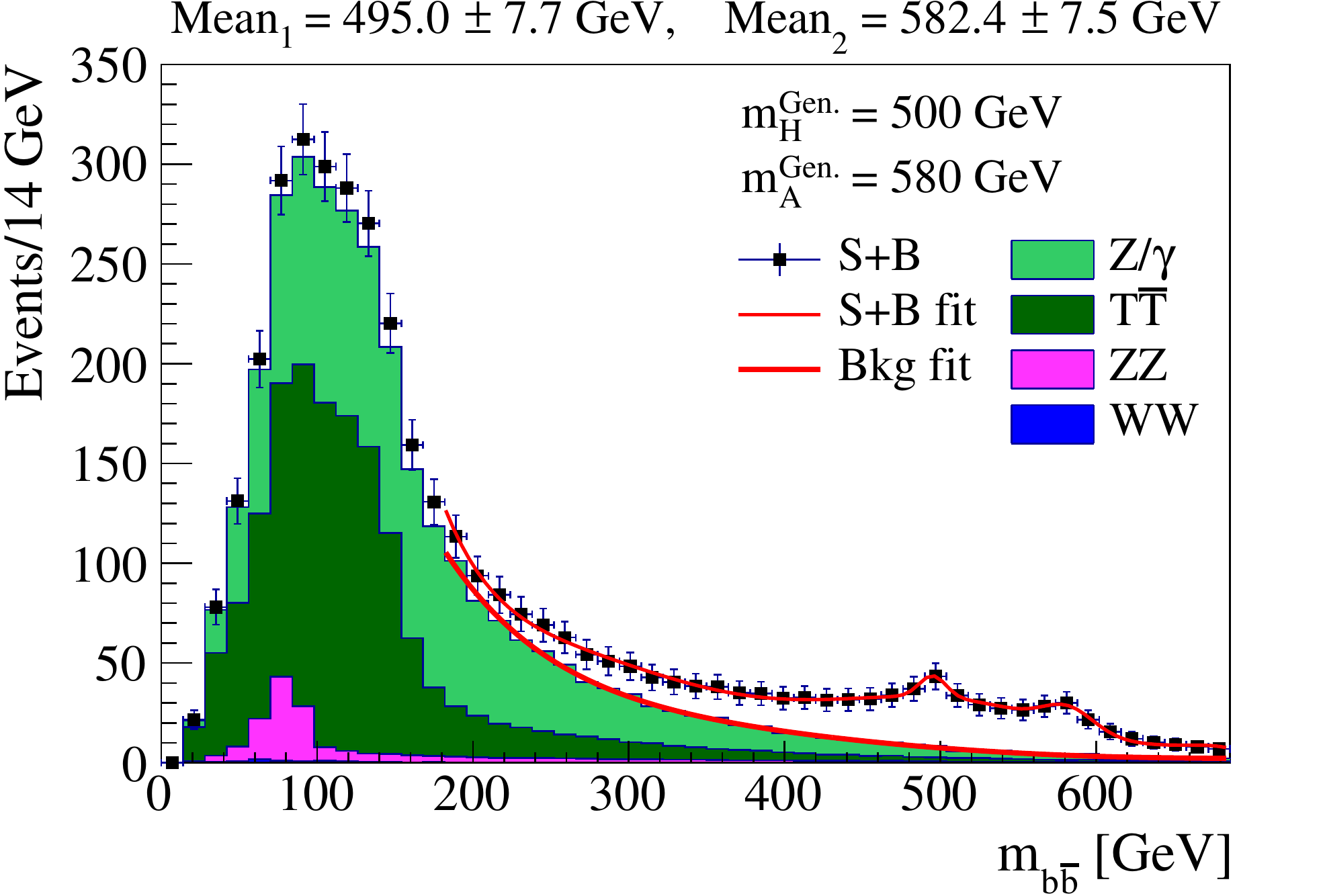}
    \caption{}
    \label{BP5}
    \end{subfigure}
\caption{Distribution of the invariant masses of the identified $b\bar{b}$ pairs in different benchmark scenarios with associated errors. Signal plus total background (S+B) fit, total background (B) fit and values of the ``Mean'' parameters are also shown. }
\label{fit}
\end{figure*}

After rescaling the four-momenta, the $b$-jets are sorted in terms of their new energies and pairing is performed as explained. Computing the invariant masses of the selected pairs, invariant mass distributions of Fig. \ref{fit} are obtained.
As seen, contributions of different background processes are shown separately and the signal contribution can be seen as a significant excess of data on top the total SM background. The $t\bar{t}$ and $Z/\gamma$ processes contribute the most to the total background and are, however, well under control. Normalization of the distributions is based on $L\times \sigma\times \epsilon$, where $L$ is the integrated luminosity which is set to 500 $fb^{-1}$ for all the scenarios, $\sigma$ is the cross section which is obtained from the total cross sections provided in Tab. \ref{BPs} and the branching fractions of the $A/H\rightarrow b\bar{b}$ decays, and $\epsilon$ is the selection efficiency which is obtained by computing the average number of reconstructed Higgs bosons in an event. 

In order to determine the reconstructed masses of the Higgs bosons, a proper fit function is fitted to the mass distributions. Fitting is performed by ROOT 5.34 \cite{root2}. The fit function employed for the total background (B) fit is a polynomial function and the fit function for the signal plus total background (S+B) fit is the combination of a polynomial function and two Gaussian functions. The two Gaussian functions are supposed to cover the signal peaks. The polynomial is first fitted to the total background distribution and the resultant fit parameters are then used as input for the S+B fit. Both B and S+B fit results are shown in Fig. \ref{fit}. As expected, fitted curves show two distinct peaks by which the reconstructed masses of the Higgs bosons $H$ and $A$ can be determined. Each Gaussian function has a ``Mean'' parameter which shows the location of the center of its associated peak. Values of the ``Mean'' parameters of the two Gaussian functions are shown in Fig. \ref{fit}. Considering the ``Mean'' parameter value as the Higgs boson reconstructed mass, reconstructed masses of the $H$ and $A$ Higgs bosons are obtained as provided in Tab. \ref{recmass}. 
\begin{table}[h]
\normalsize
\fontsize{11}{7.2} 
    \begin{center}
         \begin{tabular}{ >{\centering\arraybackslash}m{.18in} >{\centering\arraybackslash}m{.27in}  >{\centering\arraybackslash}m{.5in}  >{\centering\arraybackslash}m{.9in} >{\centering\arraybackslash}m{.9in} >{\centering\arraybackslash}m{0in} } 
\Xhline{3\arrayrulewidth} 
  &  & $m_{\,Gen.}$ & $m_{\, Rec.}$ & $m_{\,\, Corr. \,\, rec.}$ \parbox{0pt}{\rule{0pt}{1ex+\baselineskip}}\\ \Xhline{3\arrayrulewidth} 
   &   \cellcolor{blizzardblue}{BP1} & 150 & 137.2$\pm$1.9 & 146.7$\pm$5.3 \parbox{0pt}{\rule{0pt}{1ex+\baselineskip}}\\ 
  &   \cellcolor{blizzardblue}{BP2} & 200 & 188.2$\pm$1.5 & 197.7$\pm$4.9 \parbox{0pt}{\rule{0pt}{1ex+\baselineskip}}\\ 
\multirow{2}{*}[3.8pt]{\textbf{H}} & \cellcolor{blizzardblue}{BP3} & 300 & 288.8$\pm$2.4 & 298.3$\pm$5.8 \parbox{0pt}{\rule{0pt}{1ex+\baselineskip}}\\
&   \cellcolor{blizzardblue}{BP4} & 400 & 393.4$\pm$3.6 & 402.9$\pm$7.0 \parbox{0pt}{\rule{0pt}{1ex+\baselineskip}}\\
&   \cellcolor{blizzardblue}{BP5} & 500 & 495.0$\pm$7.7 & 504.5$\pm$11.1 \parbox{0pt}{\rule{0pt}{1ex+\baselineskip}}\\ \Xhline{2\arrayrulewidth}
 &   \cellcolor{blizzardblue}{BP1} & 230 & 212.0$\pm$1.5 & 220.1$\pm$4.6 \parbox{0pt}{\rule{0pt}{1ex+\baselineskip}}\\ 
  &   \cellcolor{blizzardblue}{BP2} & 280 & 268.9$\pm$1.3  & 277.0$\pm$4.4 \parbox{0pt}{\rule{0pt}{1ex+\baselineskip}}\\ 
 \multirow{2}{*}[5pt]{\textbf{A}} & \cellcolor{blizzardblue}{BP3} & 380 & 371.2$\pm$1.9 & 379.3$\pm$5.0 \parbox{0pt}{\rule{0pt}{1ex+\baselineskip}}\\
&   \cellcolor{blizzardblue}{BP4} & 480 & 475.1$\pm$3.4 & 483.2$\pm$6.5 \parbox{0pt}{\rule{0pt}{1ex+\baselineskip}}\\
&   \cellcolor{blizzardblue}{BP5} & 580 & 582.4$\pm$7.5 & 590.5$\pm$10.6 \parbox{0pt}{\rule{0pt}{1ex+\baselineskip}}\\ \Xhline{3\arrayrulewidth}
        \end{tabular} 
  \end{center}  
\caption{Generated mass ($m_{\,Gen.}$), reconstructed mass ($m_{\,Rec.}$) and corrected reconstructed mass ($m_{\,\,Corr.\,\, rec.}$) of the Higgs bosons $H$ and $A$ with associated uncertainties. Mass values are in GeV unit.}
\label{recmass} 
\end{table} 
Comparing the generated ($m_{\,Gen.}$) and reconstructed ($m_{\,Rec.}$) masses, a difference is seen between them. Such errors can be due to the uncertainties arising from the jet reconstruction algorithm, $b$-tagging algorithm, fitting method and fit function, errors in energy and momentum of the particles, etc. A thorough optimization of the jet clustering algorithm, $b$-tagging algorithm, fitting method, etc., may reduce the errors. However, since such corrections are beyond the scope of this paper, a simple off-set correction is applied to reduce errors in this study as follows. On average, the reconstructed masses of the Higgs bosons $H$ and $A$ are 9.48 and 8.08 GeV smaller than the corresponding generated masses as seen in Tab. \ref{recmass}. Hence, to reduce the errors, the reconstructed masses of the Higgs bosons $H$ and $A$ are increased by the same values. Applying the off-set correction, obtained results are provided in Tab. \ref{recmass} as corrected reconstructed masses ($m_{\,\,Corr.\,\, rec.}$). Making a comparison, it can be seen that the obtained masses are in reasonable agreement with the generated masses and therefore, it can be concluded that mass measurement is possible for both $H$ and $A$ Higgs bosons in all the considered scenarios.

\section{Signal significance}   
Observability of the Higgs bosons is assessed by computing the signal significance for each candidate mass distribution of Fig. \ref{fit}. Computation is performed by counting the number of signal and background Higgs candidate masses in the whole mass range at the integrated luminosity of 500 $fb^{-1}$. Tab. \ref{sign} provides obtained results including total signal selection efficiency, number of signal (S) and total background (B), signal to total background ratio and signal significance. Results indicate that both of the $H$ and $A$ Higgs bosons are observable with signals exceeding $5\sigma$ in all of the considered benchmark scenarios. Consequently, the region of parameter space with mass ranges $150\leq m_H \leq 500$ GeV and $230\leq m_A \leq 580$ GeV with the mass splitting of 80 GeV between the $H$ and $A$ Higgs bosons is observable at the integrated luminosity of 500 $fb^{-1}$ and $\sqrt s=1.5$ TeV. The high obtained signal significances ensure that observability is also possible at integrated luminosities lower than 500 $fb^{-1}$.
\begin{table}[!htbp]
\normalsize
\fontsize{11}{7.2} 
    \begin{center} 
         \begin{tabular}{  >{\centering\arraybackslash}m{.9in}  >{\centering\arraybackslash}m{.35in}  >{\centering\arraybackslash}m{.35in} >{\centering\arraybackslash}m{.39in}  >{\centering\arraybackslash}m{.35in} >{\centering\arraybackslash}m{.45in} >{\centering\arraybackslash}m{.35in}}  
 \Xhline{3\arrayrulewidth}
   & {BP1} & {BP2} & {BP3} & {BP4} & {BP5} \parbox{0pt}{\rule{0pt}{1ex+\baselineskip}}\\ \Xhline{3\arrayrulewidth}  
    \cellcolor{blizzardblue}{$\epsilon_{\, Total}$} & 0.27 & 0.38 & 0.42 & 0.43 & 0.44 \parbox{0pt}{\rule{0pt}{1ex+\baselineskip}}\\ 
    \cellcolor{blizzardblue}{$S$} & 1549 & 1968 & 1666 & 1216 & 715 \parbox{0pt}{\rule{0pt}{1ex+\baselineskip}}\\ 
    \cellcolor{blizzardblue}{$B$} & \multicolumn{5}{c}{3132} \parbox{0pt}{\rule{0pt}{1ex+\baselineskip}}\\  
\cellcolor{blizzardblue}{$S/B$} & 0.49 & 0.63 & 0.53 & 0.39 & 0.23 \parbox{0pt}{\rule{0pt}{1ex+\baselineskip}}\\  
\cellcolor{blizzardblue}{$S/\sqrt{B}$} & 27.7 & 35.2 & 29.8 & 21.7 & 12.8 \parbox{0pt}{\rule{0pt}{1ex+\baselineskip}}\\  
 \cellcolor{blizzardblue}{$\mathcal{L}_{\, Int.}$ [$fb^{-1}$]} &  \multicolumn{5}{c}{500} \parbox{0pt}{\rule{0pt}{1ex+\baselineskip}}\\ 
\Xhline{3\arrayrulewidth} 
\end{tabular}
  \end{center}
\caption{Total signal selection efficiency ($\epsilon_{\, Total}$), number of signal (S) and background (B) Higgs candidates in the whole mass range after all cuts, signal to background ratio, signal significance and integrated luminosity in the considered scenarios.} 
 \label{sign}
\end{table}

\section{Conclusions}   
In this study, assuming the Type-Y (flipped) 2HDM at SM-like scenario as the theoretical framework, observability of the additional CP-even and CP-odd Higgs bosons $H$ and $A$ was investigated through the signal process $e^-e^+\rightarrow AH\rightarrow b\bar{b}b\bar{b}$ at the center-of-mass energy of 1.5 TeV at a linear collider. The signal process benefits from possible enhancements in the $A/H\rightarrow b\bar{b}$ decay channels at high values of $\tb$. Such enhancements are due to the $-\tb$ factor in the $A/H$-$d$-$\bar{d}$ vertex, where $d$ is a down-type quark. Moreover, since the $A/H$-$u$-$\bar{u}$ vertex, where $u$ is an up-type quark, depends on $\cot\beta$, the dominance of the $A/H\rightarrow b\bar{b}$ decays continues even for Higgs masses $m_{A/H}$ above the threshold of the on-shell top quark pair production. Such a unique feature provided opportunity to probe a significantly large portion of the parameter space. Considering several benchmark points in the parameter space with the $H$ mass range $150\leq m_H \leq 500$ GeV and $A/H$ mass splitting of 80 GeV at $\tb=20$, event generation was performed for each scenario independently. Simulating the detector response based on the SiD detector at the ILC, observability was investigated at the integrated luminosity of 500 $fb^{-1}$ and a Higgs candidate mass distribution was obtained for each scenario. All the obtained mass distributions show significant excess of data on top of the total SM background. Two well distinguished peaks are also seen in the distributions located near the generated masses of the $H$ and $A$ Higgs bosons. Computing the signal significance corresponding to the whole mass range for each scenario, it is concluded that both of the $H$ and $A$ Higgs bosons are observable with signals exceeding $5\sigma$ in all of the tested benchmark points. In other words, the region of parameter space with mass ranges $150\leq m_H \leq 500$ GeV and $230\leq m_A \leq 580$ GeV with the mass splitting of 80 GeV between the Higgs bosons $H$ and $A$ is observable at the integrated luminosity of 500 $fb^{-1}$ and center-of-mass energy of 1.5 TeV. The reconstructed masses of the Higgs bosons which were obtained by fitting proper functions to the mass distributions are in reasonable agreement with the generated masses and indicate that mass measurement is also possible for both the $H$ and $A$ Higgs bosons in the mentioned region of parameter space. The present analysis is expected to serve experimentalists well since both of the additional neutral 2HDM Higgs bosons can be observed with possibility of mass measurement in a significant portion of the parameter space at an easily accessible integrated luminosity.   

\section*{Acknowledgements}
We would like to thank the college of sciences at Shiraz university for providing computational facilities and maintaining the computing cluster during the research program.

%



\end{document}